\documentclass[11pt,twoside]{article}
\usepackage{newpasp}
\usepackage{epsf}
\usepackage{epsfig}
\usepackage{rotating}
\usepackage{makeidx}

\begin{document}

\markboth{Lan\c{c}on and Mouhcine}{Stochastic Fluctuations}

\title{Stochastic Fluctuations in the Spectrophotometric Properties
    of Star Clusters}

\author{Ariane Lan\c{c}on \& Mustapha Mouhcine}
\affil{Observatoire de Strasbourg (UMR 7550), 11 rue de l'Universit\'e,
    F-67\,000 Strasbourg, France}


\begin{abstract}
Integrated spectrophotometric properties of stellar systems are 
intrinsically dispersed due to the stochastic nature of the
small numbers of bright stars they contain. Among clusters, only
the most massive ones can be used individually for comparison
with the mean properties predicted by population synthesis 
calculations. The appropriate minimal masses depend, among others,
on the waveband or colour index studied and on age. 
Selected indices (near-IR CO and H$_2$O, EW(H$\alpha$)) are
used to illustrate the asymmetric nature of the probability
distribution of observable properties and their dependence
on cluster mass.
\end{abstract}




\section{Introduction}

Most population synthesis codes (those based on Monte Carlo 
simulations excepted) predict {\em mean} properties of stellar populations
as a function of fundamental model parameters such as the stellar initial mass 
function (IMF) and the star formation history (SFH). 
However, for a given model the number of stars in each area of the 
HR diagram is a statistical variable obeying Poisson
statistics. The resulting intrinsic dispersion of integrated
spectrophotometric properties is observed, both as pixel to pixel 
fluctuations in otherwise uniform objects (``surface brightness 
fluctuations", Tonry \& Schneider 1988) and as cluster to cluster 
variations among cluster samples restricted to similar SFHs 
(Ferraro et al. 1995, Girardi et al. 1995). Clusters are tempting 
targets for the tests and calibrations of population synthesis
prediction because of the coeval nature of their stars. In this
context, it is important to remember that the properties
of {\em individual} clusters are representative of the mean
properties only in the limit of large star numbers, i.e. large
cluster masses (assuming a universal IMF).
Discussing how large these masses need to be in practice is the purpose
of this paper.

All results presented here are based on the population
synthesis code { P\'egase} (Fioc \& Rocca-Volmerange 2000) and
extensions thereof. They assume solar metallicity and
a Salpeter IMF extending from 0.1 to 120\,M$_{\odot}$.

\section{Luminosity fluctuations}

\begin{figure}
\plotone{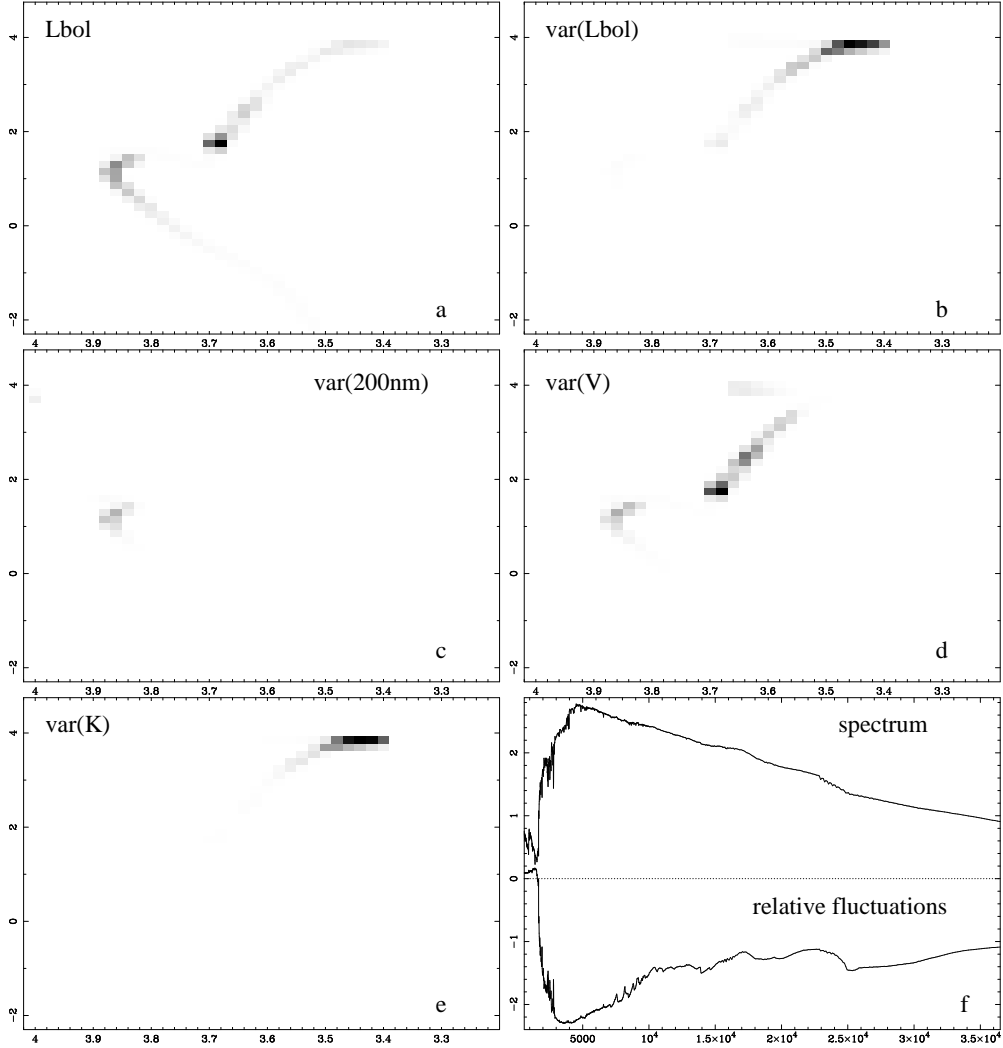}
\caption{Plots $a,b,c,d,e$ show the contribution of bins of the 
HR diagram (labelled in Log($T_{\mathrm{eff}}$) and Log($L/L_{\odot}$)) 
to $L_{\mathrm{bol}}$ and to the variance of $L_{\mathrm{bol}}$,
$L_{200\,\mathrm{nm}}$, $L_V$, $L_K$
for a solar metallicity stellar population at the age of 1\,Gyr
(Salpeter IMF, 0.1-120\,M$_{\odot}$). Post-AGB evolutionary phases that lie
outside the plotted areas are responsible for most of the UV fluctuations.
The greyscales are normalised separately in each diagram\,;
the comparison of figures $c,d$ and $e$ with $b$ shows that UV-V fluctuations 
contribute little to the variance of $L_{\mathrm{bol}}$.
Plot $f$ shows the corresponding integrated spectrum (Log($F_{\lambda}$)
in arbitrary units versus $\lambda$ in \AA ), as well as the relative
fluctuations Log($\sigma_{F_{\lambda}}/F_{\lambda}$) for a population
of $10^6$\,M$_{\odot}$ of stars.}
\label{HRplots.fig}
\end{figure}

The variance of the luminosity $L$ of a population (or of $L_{\lambda}$ to 
allow for wavelength dependence) is proportional to $\sum_i n_i L_i^2$,
where the sum extends over all luminosities $L_i$ of the HR diagram, and 
where $n_i$ is the corresponding expectation number of 
stars\footnote{In practice, the
luminosities are binned and the rms luminosity of each bin must be used\,;
the formula assumes statistical independence of the bins, which is
justified when many bins of relatively large $n_i$ but low $L_i$
contribute a negligible amount to the total luminosity or variance.}.
Intrinsically luminous stars, which already contribute significantly
to the integrated luminosity despite their relatively small numbers,
contribute even more exclusively to the variance 
(Figs.\,\ref{HRplots.fig}\,$a,b$). Figs.\,\ref{HRplots.fig}\,$c,d$ and $e$
illustrate how the strongest contributors to the flux density
depend on wavelength, using a 1\,Gyr old stellar population as 
an example. Fig.\,\ref{HRplots.fig}\,$f$ 
shows the resulting mean spectral distribution of the flux and of 
the relative rms flux deviations from this mean, $\sigma_L/L$, for a population
of 10$^6$\,M$_{\odot}$ of stars.

At all but the youngest ages, the most luminous of the red stars
determine the bolometric as well as the near-IR luminosity fluctuations.
Turn-off stars also contribute to var($L_{\mathrm{bol}}$) during the first
few $10^7$\,yrs\,; at all times
they are responsible for the optical fluctuations together
with horizontal branch and red clump stars.

The relative fluctuations around the mean flux, $\sigma_L/L$, are large
when the subpopulation that contributes most of the variance
consists of stars of large intrinsic luminosity and when the total
number of these stars is small (i.e. one is not in the large $n_i$
limit considered by Tonry \& Schneider 1988). Consider the simplified case
of a population of $N$ stars of which a mean number $\alpha N$ 
belongs to the subpopulation of interest (at the wavelength of interest).
let $l_s$ be the intrinsic individual luminosity of each subpopulation
stars, and $l_o$ that of each other star: 
$L=\alpha N l_s + (1-\alpha)N l_o = Nl$. Assuming Poisson statistics for the 
number of stars in the subpopulation ($\alpha N$), one gets
$$ \frac{\sigma_L}{L} \ = \
\frac{ \sqrt{\alpha N}\,(l_s-l_o) }{ \alpha N\,(l_s-l_o)\,+\,Nl_o }
\ = \
\frac{1}{\sqrt{N}}\,\frac{\sqrt{\alpha}\,(l_s-l_o)}{l} $$
$\alpha$, $l_s$, $l_o$ are given by population synthesis calculations. Other
assumptions (e.g. fixing other quantities than the total number of 
stars $N$) lead to slightly different formulae with similar behaviour.

\begin{table}
\small
\begin{center}
\caption[]{Minimal cluster masses, in solar units, 
ensuring that $\sigma_L/L \leq 10\,\%$.}
\vspace{1mm}

\begin{tabular}{rccc} \tableline
cluster age & $\lambda=200\,$nm & 550\,nm & 2.2\,$\mu$m \\ \tableline
\rule[0pt]{0mm}{3.7mm} 5\,Myr & $5\,10^4$ & $10^5$ & $8\,10^5$ \\
10\,Myr & $5\,10^4$ & $10^5$ & $6\,10^5$ \\
50\,Myr & $2\,10^4$ & $2\,10^4$ & $3\,10^5$ \\
200\,Myr & $10^4$ & $10^4$ & $10^6$ \\
1\,Gyr & $5\,10^5$ & $6\,10^3$ & $6\,10^5$ \\
10\,Gyr & $>\,10^6$ & $2\,10^4$ & $5\,10^5$ \\ \tableline \tableline
\end{tabular}
\end{center}
\normalsize
\end{table}

Table\,1 lists the masses required to ensure the relative luminosity 
fluctuations are below 10\,\%, with the IMF of Fig.\,\ref{HRplots.fig}.
The luminosity fluctuations directly translate into stochastic
fluctuations of the mass-to-light ratio\,($M/L$); note that differences
in the lower IMF could add significantly to the spread in $M/L$ in
cluster samples.

\section{Fluctuations in colours and spectrophotometric indices}

The statistics of flux ratios are non trivial. The 2 fluxes that define
colours or other spectrophotometric indices are in general not
independent and not Gaussian (the Gaussian approximation can be used
for the large number limit, in which case however the fluctuations
are too small to require consideration).

As a consequence of Poisson statistics, the most likely HR diagrams
for a given stellar population model underpopulate areas where the expectation
number of stars is smaller than or of the order of 1. 80\,\%
probability intervals, defined so that the statistical variable (e.g.
the number of stars of interest) has probabilities of 10\,\% to lie 
outside the interval on either side, are centered on a number smaller
than the expectation value\,; their size is not equal to
2\,$\times$\,1.28\,$\sigma$, as it would be for Gaussians. In
practice, the stars with high $L$ and small expectation numbers are red,
and the most likely colours will thus be bluer than the mean.
This explains the behaviour of the Monte-Carlo simulations of Santos
\& Frogel (1997), and in particular their Fig.\,5\,: in a population
of $10^3$ stars, the number of post main sequence stars evolves
with time from about 10 at 50\,Myr to about 100 at 1\,Gyr, 
and the expected number of luminous cool giants (red
supergiants of AGB stars) is of the order of one, resulting in most
common J-K colours significantly below the large cluster limit. 

\begin{figure}
\plotone{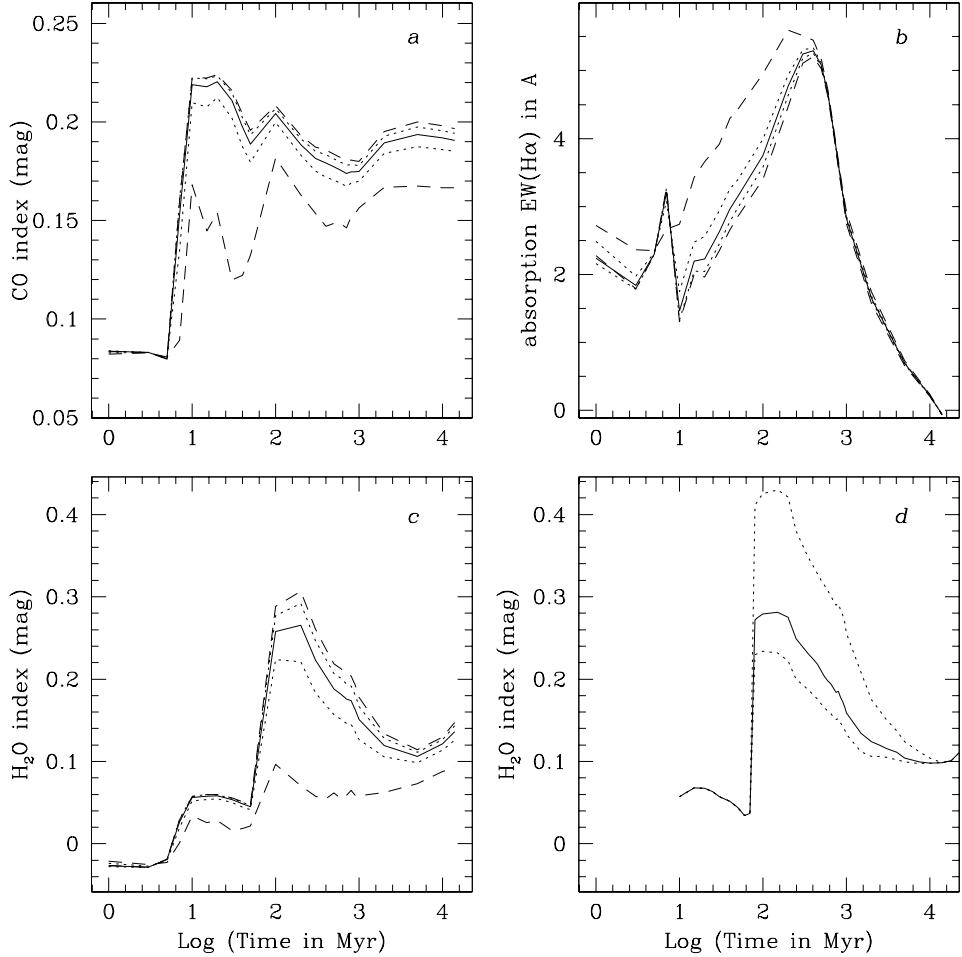}
\caption[]{In $a$, $b$ and $c$, the solid line describes the mean evolution
of selected spectrophotometric indices. The dashed and dotted lines
show the upper and lower boundaries of 80\,\% probability
intervals for these indices, for stellar populations of respectively
$10^3$ and $10^4$\,M$_{\odot}$ of stars. Plot\,$d$ compares the 
predicted mean evolutions of the H$_2$O index of plot\,$c$, with
three different assumptions for the effective temperature 
scale of variable stars of the TP-AGB\,: the dotted lines are
considered as extremes still compatible with the available 
litterature, the solid line is a currently adopted intermediate
scale (note that slightly different timesteps and stellar libraries
were used for plots $c$ and $d$). Here, all TP-AGB stars are assumed
to be oxygen rich.}
\label{indices.fig}
\end{figure}

When the baseline between the two passbands defining a 
spectrophotometric index is small, the corresponding fluxes in 
most cases originate from the same population and can be 
considered 100\,\% correlated\footnote{Exceptions
are indices such as the strenth of optical or near-IR emission lines,
for which the line flux is dominated, and the continuum contaminated,
by recombination radiation due to the presence of hot stars.}.
Fig.\,\ref{indices.fig} shows the 80\,\% probability limits obtained with
this assumption for 3 commonly used indices (gas recombination is
radiation not included). As expected, the behaviour of the  probability
intervals is complex. The stochastic fluctuations of the CO index
for instance are largest when red supergiants exist, because the latter
have particularly strong CO bands\,; at later stages, fluctuations
in the numbers of the current most luminous red stars matter less,
since red giants of various ages and luminosities have relatively similar
CO bands.

The mass required for a meaningful direct comparison between 
predicted mean properties and those of a real stellar population
depends on the scientific application. H$_2$O bands are strongest in
the stars of the upper AGB (TP-AGB), which are most important between 
$10^8$ and a few $10^9$\,yrs. These stars are still poorly understood.
If the purpose of a cluster observation is to test the effective
temperature scale of TP-AGB star spectra (e.g. Lan\c{c}on et al. 1999),
Fig.\,\ref{indices.fig}\,$d$ shows that $10^4$\,M$_{\odot}$ of stars
(i.e. typical LMC cluster masses) 
are only marginally sufficient. More appropriate clusters should have 
$10^5$\,M$_{\odot}$ or more. 

\section{Conclusions}

Stochastic fluctuations due to small numbers of bright stars need to
be considered when stellar populations are compared to population
synthesis models, be it using star counts or using integrated properties.
The most probable spectrophotometric properties of small clusters
are usually different from their expectation value (i.e. the properties in
the large cluster limit), leading to systematic effects in the determination
of age, metallicity or other fundamental parameters. The adequate
definition of a {\em massive cluster}, for which these effects would
be negligible, depends strongly on the spectrophotometric property
studied and on the star formation history. The cluster populations
formed in galaxy mergers, thoroughly discussed during this workshop
and known to contain objects of more than
10$^6$\,M$_{\odot}$, are becoming accessible to spectrographs on
large telescopes and clearly represent  important targets for 
population synthesis studies of the near future (e.g. Mouhcine \&
Lan\c{c}on, this volume).

\acknowledgments
We thank J.L.~Vergely and D.~Kunth for motivating discussions on
stochastic fluctuations and their consequences.

\vspace{-3mm}

\begin{question}{J.\,Gallagher}
WR stars being intrinsically rare objects, how can there be so many WR 
clusters\,?
\end{question}
\begin{answer}{A.\,Lan\c{c}on}
Actually, WR stars aren't that rare, at least at solar metallicity. 
According to Schaerer \& Conti (1998, ApJ, 497, 618), their numbers are of the  
same order as those of O stars over significant starburst age 
ranges and, with a solar neighbourhood initial mass function, 
one finds about one O star for every 10$^3$\,M${_\odot}$
of newly formed stars (e.g. Leitherer \& Heckman 1995, ApJS, 96, 9).
Stochastic fluctuations must be considered in individual young
clusters of less than $\sim 10^5$\,M$_{\odot}$, in particular 
when measuring ratios of different types of WR stars. They
will average out over the many clusters of a WR galaxy. 
\end{answer}

\begin{question}{S.\,Portegies Zwart}
Current dynamical simulations of clusters usually don't exceed 10$^4$ stars. 
Are their predictions useless\,?
\end{question}
\begin{answer}{A.\,Lan\c{c}on}
The good thing about numerical simulations is that they keep
you aware of the number of stars you are dealing with, a point
that one easily forgets when looking at the integrated photometric
properties of a population from a distance! Maybe one should focus
on the properties that don't depend so much on small numbers of 
bright stars first; by the time we will have tested these predictions
thoroughly, it is likely that computers will have improved enough
to increase the sizes of simulations...
\end{answer}

\end{document}